\documentclass[12pt,a4paper]{article}

\pdfoutput=1

\usepackage[DIV13,BCOR0mm]{typearea}

\usepackage{wasysym}

\usepackage{graphicx}

\newcommand{\beq}{\begin{equation}}
\newcommand{\eeq}{\end{equation}}
\newcommand{\beqa}{\begin{eqnarray}}
\newcommand{\eeqa}{\end{eqnarray}}
\newcommand{\beqan}{\begin{eqnarray*}}
\newcommand{\eeqan}{\end{eqnarray*}}

\newcommand{\bigfrac}[2]{\mbox{$\displaystyle\frac{#1}{#2}$}}

\newcommand{\D}{\mathrm{d}}

\newcommand{\grad}{\mbox{grad}\,}
\newcommand{\divg}{\mbox{div}\,}

\newcommand{\tr}{\mbox{tr}\,}

\newcommand{\tra}{^{\mathrm{T}}}    
\newcommand{\dev}{^{\mathrm{D}}}    

\newcommand{\pabl}[2]{\frac{\partial #1}{\partial #2}}

\newcommand{\nl}{\nonumber\\}
\newcommand{\degC}{\ensuremath{^\circ\mathrm{C}}}    
\newcommand{\degN}{\ensuremath{^\circ\mathrm{N}}}    
\newcommand{\degS}{\ensuremath{^\circ\mathrm{S}}}    
\newcommand{\degE}{\ensuremath{^\circ\mathrm{E}}}    
\newcommand{\degW}{\ensuremath{^\circ\mathrm{W}}}    

\def\pmb#1{\setbox0=\hbox{#1}%
  \kern-.025em\copy0\kern-\wd0
  \kern.05em\copy0\kern-\wd0
  \kern-.025em\raise.0433em\box0}

\def\pmbm#1{\setbox0=\hbox{$#1$}%
  \kern-.025em\copy0\kern-\wd0
  \kern.05em\copy0\kern-\wd0
  \kern-.025em\raise.0433em\box0}

\def\vecbi#1{\relax\ifmmode\mathchoice
  {\mbox{\boldmath$\relax\displaystyle#1$}}
  {\mbox{\boldmath$\relax\textstyle#1$}}
  {\mbox{\boldmath$\relax\scriptstyle#1$}}
  {\mbox{\boldmath$\relax\scriptscriptstyle#1$}}\else
  \hbox{\boldmath$\relax\textstyle#1$}\fi} 

\def\vecbu#1{\relax\ifmmode\mathchoice
  {\mbox{\boldmath$\bf\displaystyle#1$}}
  {\mbox{\boldmath$\bf\textstyle#1$}}
  {\mbox{\boldmath$\bf\scriptstyle#1$}}
  {\mbox{\boldmath$\bf\scriptscriptstyle#1$}}\else
  \hbox{\boldmath$\bf\textstyle#1$}\fi}    

\def\tenssu#1{\relax\ifmmode\mathchoice
    {\mbox{$\sf\displaystyle#1$}}
    {\mbox{$\sf\textstyle#1$}}
    {\mbox{$\sf\scriptstyle#1$}}
    {\mbox{$\sf\scriptscriptstyle#1$}}\else
    \hbox{$\sf\textstyle#1$}\fi}           

\begin{document}

\title{\textbf{Fluid dynamics of planetary ices}}

\author{\textsc{Ralf Greve}\thanks{E-mail: greve@lowtem.hokudai.ac.jp}\\[0.5ex]
        {\normalsize Institute of Low Temperature Science, Hokkaido University,}\\[-0.25ex]
        {\normalsize Kita-19, Nishi-8, Kita-ku, Sapporo 060-0819, Japan}}

\date{}

\maketitle

\begin{abstract}
The role of water ice in the solar system is reviewed from a
fluid-dynamical point of view. On Earth and Mars, water ice
forms ice sheets, ice caps and glaciers at the surface, which
show glacial flow under their own weight. By contrast, water
ice is a major constituent of the bulk volume of the icy
satellites in the outer solar system, and ice flow can occur as
thermal convection. The rheology of polycrystalline aggregates
of ordinary, hexagonal ice~Ih is described by a power law,
different forms of which are discussed. The temperature
dependence of the ice viscosity follows an Arrhenius law.
Therefore, the flow of ice in a planetary environment
constitutes a thermo-mechanically coupled problem; its model
equations are obtained by inserting the flow law and the
thermodynamic material equations in the balance laws of mass,
momentum and energy. As an example of gravity-driven flow, the
polar caps of Mars are discussed. For the north-polar cap,
large-scale flow velocities of the order of
$0.1\ldots{}1\,\mathrm{mm\,a^{-1}}$ are likely, locally
enhanced by a factor ten or more in the vicinity of surface
scarps/troughs. By contrast, the colder south-polar cap is
expected to be almost stagnant. Tidally heated convection is
discussed for the example of the icy crust of Europa, where a
two-dimensional model predicts the formation of an upper,
conductive lid and a lower, convective layer with flow
velocities of the order of $100\,\mathrm{mm\,a^{-1}}$. Very
little is known about the fluid-dynamical relevance of
high-pressure phases of water ice as well as ices made up of
other materials.
\end{abstract}

\vspace*{-176mm}

\noindent{\footnotesize\emph{GAMM-Mitteilungen} \textbf{29}(1),
                       29--51 (2006)
\\[-0.3ex]
http://www.interscience.wiley.com/
\\
--- Author's version ---}

\vspace*{161mm}

\section{Introduction}
\label{sec_intro}

Water ice is an abundant material in the solar system. In the inner
solar system, it plays an important role on Earth and on Mars, where
it forms a cryosphere as an active, dynamic part of the respective
climate system. Approximately 10\% of the land surface of the
present-day Earth, or $14.6\times{}10^6\,\mathrm{km^2}$, are covered
by ice sheets, ice caps and glaciers, the total volume of which is
approximately $28.7\times{}10^6\,\mathrm{km^3}$ \cite{church_etal_01}.
By far the largest single ice body is the Antarctic ice sheet, which
alone contains about 90\% of this ice. All of these terrestrial ice
bodies are subject to gravity-driven glacial flow with typical
velocities of tens to hundreds of meters per year.  Further components
of the terrestrial cryosphere are the floating ice shelves and sea
ice, seasonal snow and ground ice (permafrost).

The ice sheets, ice shelves, ice caps and glaciers have formed by
accumulated snowfall over centuries, millenia and more. Over these
time-scales, the Earth's climate has experienced significant changes
known as glacial-interglacial cycles, which are driven by periodic
changes of the orbital parameters obliquity (axial tilt), eccentricity
and precession (``Milankovitch cycles''). The last 800,000 years have
been characterized by a strong dominance of the 100,000-year
eccentricity cycle \cite{raymo_etal_97}, with a sequence of shorter
interglacials (warm periods) like the current Holocene and longer
glacials (ice ages). At the last glacial maximum approximately 20,000
years ago, large parts of north America and Eurasia were covered by
ice sheets which no longer exist, and the global ice volume was about
three times larger than at present.

On Mars, the polar ice caps are one of the most prominent surface
features. The seasonal caps, which can extend down to latitudes of
approximately $55\degN$/S, consist of only some ten centimeters of
CO$_2$ snow which sublimes into the atmosphere during the respective
spring season. By contrast, the smaller residual caps poleward of
approximately $80\degN$/S survive the summer seasons, and they are
underlain by massive topographic structures, which are known as the
polar layered deposits \cite{thomas_etal_92}. The residual caps and
the underlying layered deposits are considered to be geomorphological
units and shall be referred to as the north- and south-polar cap
(NPC/SPC), respectively.  The Mars Orbiter Laser Altimeter (MOLA)
measurements of the Mars Global Surveyor (MGS) spacecraft have
provided a very precise mapping of the surface topographies of the
polar caps \cite{smith_etal_99a, zuber_etal_98}.  Combined with the
estimated cap margins and equilibrated ground topographies discussed
later (Sect.~\ref{sect_mpc_sico}), this yields for the NPC a volume of
about $1.2\times{}10^6\,\mathrm{km^3}$ and an area of
$1.1\times{}10^6\,\mathrm{km^2}$, and for the SPC a volume of
$1.8\times{}10^6\,\mathrm{km^3}$ and an area of
$1.7\times{}10^6\,\mathrm{km^2}$. Due to isostatic deflection of the
underlying lithosphere, the real volumes may be up to 30\% larger.  In
any case, the NPC and SPC are the largest known water reservoirs on
Mars. Further constituents may be dust, CO$_2$ ice and CO$_2$
clathrate hydrate.  Comparable or even larger amounts of water may be
stored as permafrost in the ground \cite{clifford_etal_00}; an idea
which was corroborated by the spectrometric detection of mid- and
high-latitude subsurface layers enriched in hydrogen, interpreted as
ground ice \cite{boynton_etal_02}.

Similar to the situation on Earth, the Martian polar caps are active
components of the climate system which interact with the atmosphere
thermally, orographically and by condensation and sublimation
processes of water vapour. Their present topographies are the result
of the climatic history over at least the last millions of years,
which were probably characterized by climate cycles as a consequence
of strong, quasi-periodic variations of the orbital parameters
obliquity, eccentricity and precession on time-scales of
$10^5$--$10^6$ years \cite{laskar_etal_04}. This idea is supported by
the light-dark layered deposits of the polar caps, which are exposed
in the scarps and troughs in the ice surface and close to the margins,
and which indicate a strongly varying dust content of the ice due to
varying atmospheric conditions in the past.
Further, Head et~al.\ \cite{head_etal_03} suggested that Mars
underwent ``ice ages'' during periods of high obliquity
like that from about 2.1 to 0.4 million years ago (with obliquity
maxima of ${}\approx{}35^\circ$). These ice ages are
supposedly characterized by warmer polar climates, enhanced mass loss
of the polar caps due to sublimation and the formation of meters-thick
ice deposits equatorward to approximately $30\degN$/S.

For the Moon and Mercury, which are both devoid of any significant
atmosphere and therefore subject to intensive solar radiation,
water ice may be cold-trapped inside permanently shadowed craters
at the poles, which is possible due to the very small tilts of the
rotational axes. Some evidence for this idea was provided by the
space-probe Lunar Prospector, which detected hydrogen in the
vicinity of the lunar poles by neutron spectrometry
\cite{feldman_etal_01}, and, for the case of Mercury, by
terrestrial radar mapping which revealed a highly reflective
region on the north pole \cite{slade_etal_92}.

In the outer solar system (beyond the asteroid belt), owing to the
very low temperatures, evaporation of water ice at the surface of
planetary bodies becomes so small that this substance survives for
time-spans comparable to the age of the solar system
\cite{watson_etal_63}. Therefore, while in the inner solar system
the geological evolution of the planets is dominated by rock
material, the history of a large number of bodies in the outer
solar system is dominated by water ice and other moderately
volatile substances. Of course, this statement does not hold for
the giant planets, which do not have a solid surface and are made
up of highly volatile substances, mainly hydrogen and helium.
Thus, it refers to the large moons of the Jovian, Saturnian,
Uranian and Neptunian systems, Pluto and smaller moons, asteroids
and planetary ring particles. Of particular interest are the
Jovian moon Europa, for which there is strong evidence for an
outer icy shell underlain by a deep ocean \cite{kargel_etal_00},
and the Saturnian moon Titan, for which the findings of the
recently landed space-probe Huygens suggest a solid surface made
up mainly of water ice and a methane hydrosphere with rainfall,
rivers and seas (see http://huygens.esa.int).

In this review paper, the focus will be on \emph{flowing} water ice,
which can be driven either by gravity forces or by thermal convection.
The layout is as follows. In Sect.~\ref{sect_ice_rheology}, the
rheology of polycrystalline aggregates of water ice is discussed,
including the effects of different creep mechanisms, partial melting
and impurities (dust), and suitable flow laws (stress-strain-rate
relations) are given.  This allows the formulation of a closed set of
model equations for the thermo-mechanically coupled problem of ice
flow, which is done in Sect.~\ref{sect_model_ice_flow}. Applications
to the Martian polar caps are presented in Sect.~\ref{sect_mpc}. New
results are shown for their large-scale dynamics under assumed
steady-state conditions (Sect.~\ref{sect_mpc_sico}), and the local
influence of the prominent scarps and troughs in the surface is
treated (Sect.~\ref{sect_mpc_scarp}). Simulations on tidally heated
convection in the icy shell of Europa are discussed in
Sect.~\ref{sect_eur_is}. Finally, Sect.~\ref{sect_conclusion}
concludes the paper.

\section{Ice rheology}
\label{sect_ice_rheology}

\subsection{Structure of ice}
\label{sect_ice_structure}

\begin{figure}[htb]
  \centering
  \includegraphics[scale=0.8]{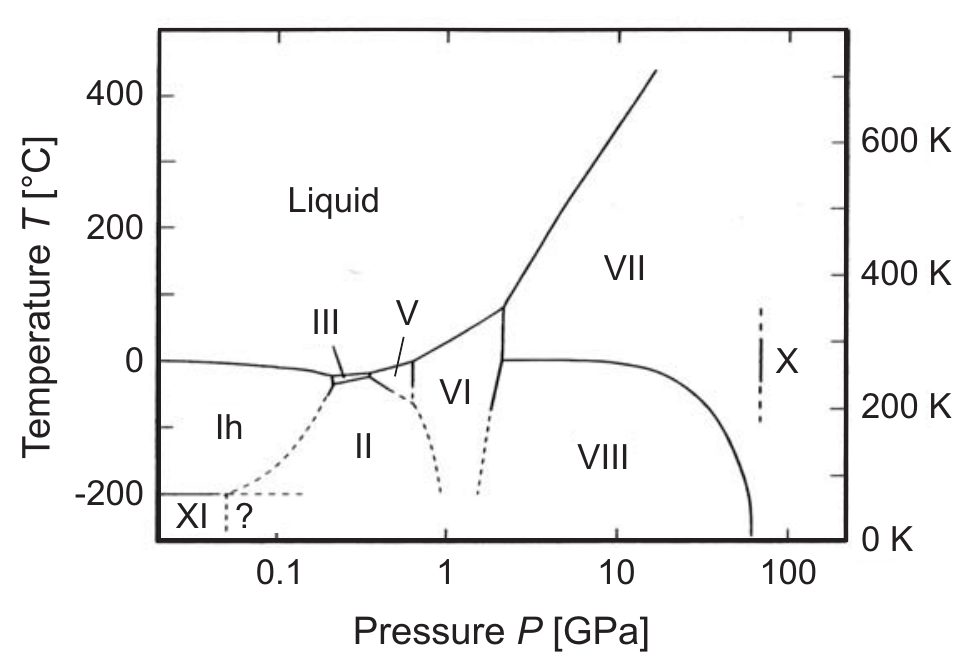}
  \caption{Phase diagram of the ice-water system.
  Only stable phases are shown.
  Figure by Petrenko and Whitworth
  \cite[their Fig.~11.2 on p.~253]{petrenko_whitworth_99}.}
  \label{fig_ice_phases}
\end{figure}

Water ice can exist in a great variety of different phases
(Fig.~\ref{fig_ice_phases}, Ref.~\cite{petrenko_whitworth_99}). Most
of these phases form under high pressure, which results in a
denser packing of the water molecules compared to the ``ordinary''
ice~Ih. The stability range of the latter is for pressures
$P\apprle{}200\;\mathrm{MPa}$, which is equivalent to the
hydrostatic pressure of an ice layer (density
$\rho_\mathrm{i}=910\,\mathrm{kg\,m^{-3}}$) of approximately 22~km
on Earth (gravity acceleration $g=9.81\,\mathrm{m\,s^{-2}}$),
60~km on Mars ($g=3.72\,\mathrm{m\,s^{-2}}$) and 165~km on Europa
($g=1.32\,\mathrm{m\,s^{-2}}$) and Titan
($g=1.35\,\mathrm{m\,s^{-2}}$). We will limit our discussions to
the outer shells of the planetary bodies, so that only ice~Ih
needs to be considered in the following.

\begin{figure}[htb]
  \centering
  \includegraphics[scale=0.92]{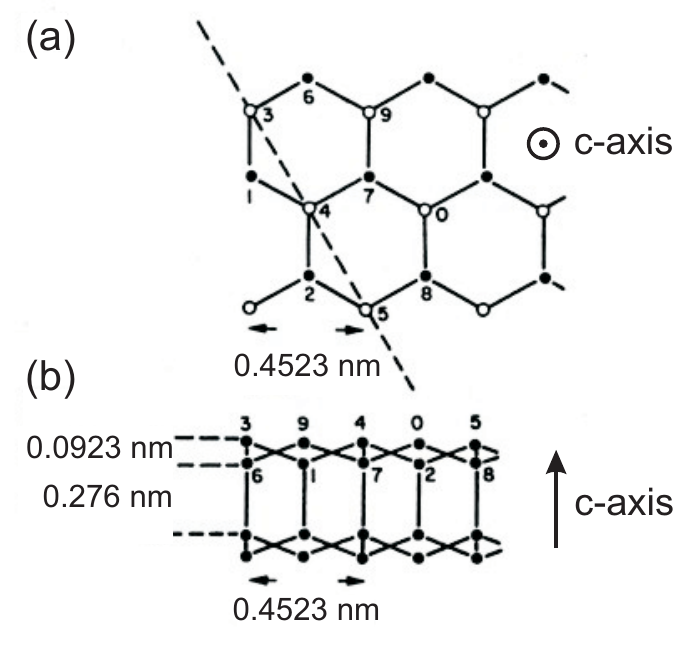}
  \caption{Structure of an ice crystal. The circles denote the oxygen
  atoms of the H$_2$O molecules. (a) Projection on the basal plane.
  (b) Projection on plane indicated by the broken line in (a).
  Figure by Paterson \cite[his Fig.~5.1 on p.~80]{paterson_94}.}
  \label{fig_iceIh}
\end{figure}

Ice~Ih forms hexagonal crystals, that is, the water molecules are
arranged in layers of hexagonal rings
(Fig.~\ref{fig_iceIh}, Ref.~\cite{paterson_94}).
The plane of such a layer is called the basal
plane, which actually consists of two planes shifted slightly (by
0.0923~nm) against each other. The direction perpendicular to the
basal planes is the optic axis or c-axis, and the distance between
two subsequent basal planes is 0.276~nm. This leads to the very
low packing factor of 34\%, which is responsible for the density
anomaly of ice~Ih (smaller density than liquid water). Further, the
basal planes can glide on each other when a shear stress is
applied, comparable to the deformation of a deck of cards. This
effect is strongly enhanced by the existence and generation of
dislocations (structural defects) in real crystals, and the
mechanism is consequently called \emph{dislocation creep}.

On the macro-scale, ice aggregates on planetary bodies (grounded
ice sheet/cap/gla\-cier, ground ice, floating ice shell etc.) are
composed of a vast number of individual crystals. For instance,
for terrestrial ice sheets and glaciers, the typical grain size is
of the order of millimeters to centimeters. Such a compound is
called \emph{polycrystalline ice}. At the time of formation, it
can be assumed that the orientation of the crystals is completely
at random, so that the macroscopic behaviour of the compound will
be isotropic. Ice-core studies on Earth have revealed that in the
course of time anisotropic fabrics can develop due to the strain
history which a piece of ice experiences during its motion in an
ice sheet or glacier (e.g.\ \cite{azuma_etal_00,
thorsteinsson_96}; see also \cite[this volume]{placidi_etal_06}).
However, since essentially nothing is known about the fabric of
extraterrestrial ice bodies, we will assume isotropic conditions
for simplicity in the following.

\subsection{Pressure melting point}
\label{sect_pressure_melting}

\begin{figure}[htb]
  \centering
  \includegraphics[scale=0.8]{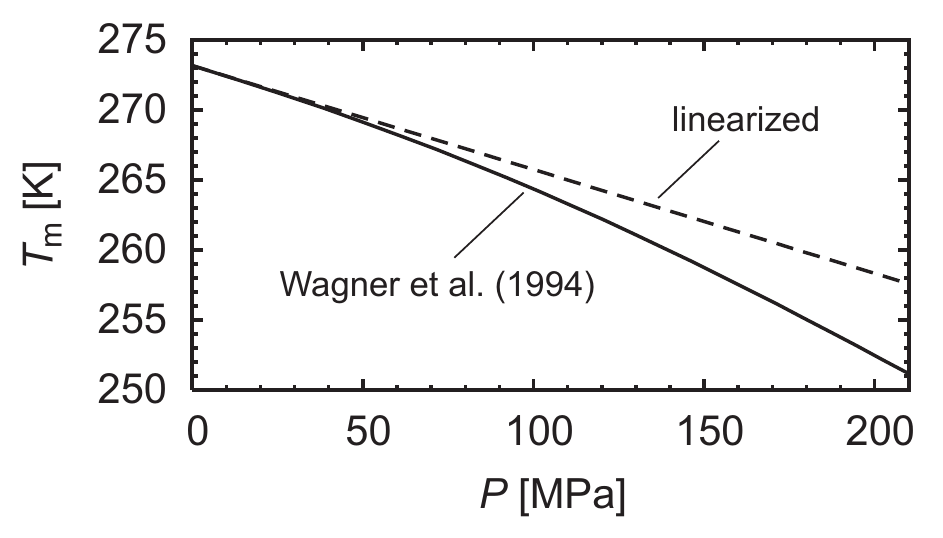}
  \caption{Pressure melting point of ice~Ih by
  Wagner et~al.\ \cite{wagner_etal_94}
  [Eq.~(\ref{eq_melt_wagner}), solid line] and linearized
  relation (\ref{eq_melt_linear}) (dashed line).}
  \label{fig_ice_melt}
\end{figure}

The melting temperature of pure ice~Ih depends on the pressure
$P$. It can therefore be written as
\beq
  T_\mathrm{m} = T_0 - f(P),
  \label{eq_p_melt_temp}
\eeq
where $T_0=273.16\;\mathrm{K}$ is the melting temperature for
$P_0=611.657\,\mathrm{Pa}$, that is, at the triple point of water.
According to \cite{wagner_etal_94}, the melting-point depression
$f(P)$ is in implicit form
\beq
  \begin{array}{c}
    \bigfrac{P}{P_0}
    = 1-\mbox{626,000}\,(1-\Theta^{-3})+\mbox{197,135}\,(1-\Theta^{21.2}),
    \\[2ex]
    \Theta = \bigfrac{T_\mathrm{m}}{T_0} = \bigfrac{T_0-f(P)}{T_0},
  \end{array}
  \label{eq_melt_wagner}
\eeq
where $P_0\le{}P\le{}209.9\,\mathrm{MPa}$ and
$T_0\ge{}T_\mathrm{m}\ge{}251.165\,\mathrm{K}$
(Fig.~\ref{fig_ice_melt}). For moderate pressures,
Eq.~(\ref{eq_melt_wagner}) can be linearized as
\beq
  f(P) = \beta{}P,
  \label{eq_melt_linear}
\eeq
where $\beta=7.42\times{}10^{-8}\,\mathrm{K\,Pa^{-1}}$ is the
Clausius-Clapeyron constant for pure ice \cite{paterson_94}. The
error of this linearization is negligible for $P\le{}10\,\mathrm{MPa}$,
less than 10\% for $P\le{}50\,\mathrm{MPa}$ and approximately 30\% for
$P\approx{}200\,\mathrm{MPa}$ (Fig.~\ref{fig_ice_melt}).

Impurities of different kinds alter the melting point further. For
terrestrial, air-saturated glacier ice, Paterson \cite{paterson_94}
reports a modified Clausius-Clapeyron constant of
$\beta=9.8\times{}10^{-8}\,\mathrm{K\,Pa^{-1}}$ for the linearized
melting-point depression (\ref{eq_melt_linear}). If salts are
present, an additional melting-point depression occurs which
depends on the type of salts and their concentration.
Intermixtures of ammonia, which may play a role in a supposed
subsurface ocean on Titan below an outer ice-Ih layer
\cite[and references therein]{sohl_etal_03}, have a similar effect.

\subsection{Flow of polycrystalline ice}
\label{sect_ice_flow}

\begin{figure}[htb]
  \centering
  \includegraphics[scale=0.9]{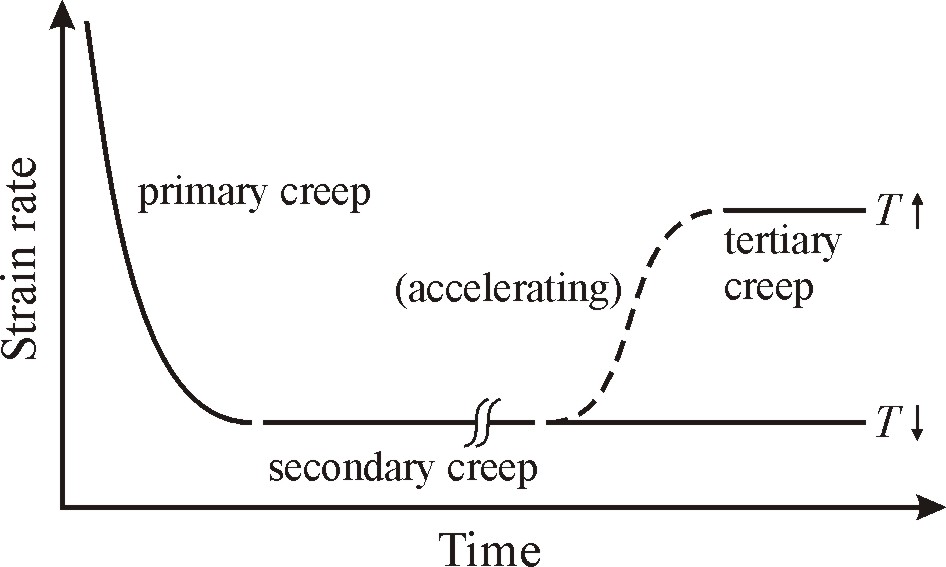}
  \caption{Creep response of a sample of polycrystalline ice to a
  constant stress for high (${}\!\apprge\!\!{}-10\degC$, marked by
  $T\!\uparrow$) and low (${}\!\apprle\!\!{}-10\degC$, marked by
  $T\!\downarrow$)
  temperatures.}
  \label{fig_ice_creep}
\end{figure}

When a specimen of polycrystalline ice is subjected to a constant
normal or shear stress, it responds with a permanent deformation,
which continues as long as the stress is applied. Typically, an
initial, instantaneous elastic deformation of the polycrystalline
aggregate is followed by a phase called primary creep during which
the strain rate decreases continuously. This behaviour is related
to the increasing geometric incompatibilities of the deforming
single crystals with different orientations. After some time, a
minimum strain rate is reached which remains constant in the
following, so that the strain increases linearly with time. This
phase is known as secondary creep. Especially in case of high
temperatures ($\apprge{}-10\degC$), at a later stage dynamic
recrystallisation (nucleation and growth of crystals which are
favourably oriented for deformation) sets in, which leads to
accelerated creep and finally a constant strain rate significantly
larger than that of the secondary creep. This is called tertiary
creep (Fig.~\ref{fig_ice_creep}, Ref.~\cite{paterson_94}).
Therefore, it can be assumed in good approximation that in
deforming ice masses secondary creep prevails for low temperatures
($\apprle{}-10\degC$), whereas tertiary creep prevails for high
temperatures ($\apprge{}-10\degC$), so that the strain rate can be
expressed as a unique function of the stress, the ice temperature
and the pressure.

On the basis of these considerations, several forms of a
non-linear viscous rheology for the flow of polycrystalline ice
for different stress, strain-rate and temperature regimes have
been proposed. They have in common to relate the strain-rate
tensor $\vecbi{D}=\mbox{sym}\,\grad\vecbi{v}$ (velocity
$\vecbi{v}$) to the Cauchy stress deviator $\vecbi{t}\dev$, and
can be subsumed as
\beq
  \vecbi{D}
  = EA(T,P)\frac{\sigma^{n-1}}{d^p}\,\vecbi{t}\dev,
  \label{eq_flowlaw_general}
\eeq
where $\sigma=[\tr(\vecbi{t}\dev)^2/2]^{1/2}$ is the
effective stress, $n$ is the stress exponent, $d$ is the grain
size and $p$ is the grain-size exponent
(e.g.\ \cite{durham_etal_97a, goldsby_kohlstedt_97, greve_mahajan_05,
paterson_94, vanderveen_99}). The flow rate factor $A(T,P)$
depends via the Arrhenius law
\beq
  A(T,P)=A_0\,e^{-(Q+PV)/RT}
  \label{eq_arrhenius_general}
\eeq
on the absolute temperature $T$ and the pressure $P$, where $A_0$ is the
pre-exponential constant, $Q$ is the activation energy, $V$ is the
activation volume and $R=8.314\;\mathrm{J\;mol^{-1}\,K^{-1}}$ is the
universal gas constant. The flow enhancement factor $E$ is equal to
unity for pure ice and can deviate from unity due to the softening or
hardening effect of impurities in the ice. Since polycrystalline ice is
described as a density-preserving (incompressible) medium, the pressure
is a free field (not governed by a material equation), and the full
Cauchy stress tensor $\vecbi{t}$ is related to the traceless deviator
$\vecbi{t}\dev$ by
\beq
  \vecbi{t} = -P\,\mathbf{1} + \vecbi{t}\dev,
\eeq
where $\mathbf{1}$ denotes the unity tensor.

Since appropriate values for the activation volume $V$ are poorly
constrained and the pressure effect is very small for typical
thicknesses of ice sheets and caps, we account for it in an
approximate way by setting $V=0$ and measuring the temperature
relative to the pressure melting point $T_\mathrm{m}$ (see
Sect.~\ref{sect_pressure_melting}) instead. To this end, the
\emph{homologous temperature}
\beq
  T' = T - T_\mathrm{m} + T_0 = T + f(P)
\eeq
is introduced, and the rate factor can be simplified as
\beq
  A(T')=A_0\,e^{-Q/RT'},
  \label{eq_arrhenius}
\eeq
which depends now exclusively on $T'$ \cite{paterson_94,
vanderveen_99}. For more details on this approach see
Appendix~\ref{sect_pressure}.

The general power law (\ref{eq_flowlaw_general}) can be inverted
as follows. For the effective strain rate
$\delta=(\tr\vecbi{D}^2/2)^{1/2}$ we obtain
\beq
  \delta = EA(T')\frac{\sigma^n}{d^p},
  \label{eq_delta_sigma}
\eeq
or equivalently, solved for $\sigma$,
\beq
  \sigma = [EA(T')]^{-1/n} d^{p/n}
           \delta^{1/n}.
  \label{eq_sigma_delta}
\eeq
Inserting this in (\ref{eq_flowlaw_general}) and solving for
$\vecbi{t}\dev$ yields
\beqa
  \vecbi{t}\dev
  &=& [EA(T')]^{-1} d^p
  \left( [EA(T')]^{-1/n} d^{p/n} \delta^{1/n} \right)^{1-n}
  \vecbi{D}
  \nl
  &=& [EA(T')]^{-1/n} \frac{d^{p/n}}{\delta^{1-1/n}}\,\vecbi{D}.
\eeqa
By introducing the stress enhancement factor $E_\mathrm{s}=E^{-1/n}$ and the
associated rate factor $B(T')=[A(T')]^{-1/n}$, this can be written
as
\beq
  \vecbi{t}\dev
  = E_\mathrm{s}B(T') \frac{d^{p/n}}{\delta^{1-1/n}}\,\vecbi{D}.
  \label{eq_flowlaw_inv_general}
\eeq

For terrestrial ice, the well-established Glen's flow law [which
actually goes back to \cite{nye_57} in the general tensorial form]
uses the stress exponent $n=3$, the grain-size exponent $p=0$ and
for the temperature range $T'\le{}263\;\mathrm{K}$ the
pre-exponential constant
$A_0=3.985\times{}10^{-13}\;\mathrm{s^{-1}\,Pa^{-3}}$ and the
activation energy $Q=60\;\mathrm{kJ\;mol^{-1}}$
\cite{paterson_94}. The rheology defined by these parameters
describes the grain-size-independent flow mechanism of dislocation
creep, which prevails in terrestrial glaciers and ice sheets. The
flow enhancement factor for ice formed during glacial periods is
often set to $E=3$, interpreted as the softening influence of very
small amounts of fine dust, approximately
$1\;\mathrm{mg\;kg^{-1}}$ with particle sizes of 0.1 to
$2\,\mu\mathrm{m}$ \cite{hammer_etal_85}. This softening is
attributed to thin films of liquid water which form around the
dust particles and lubricate ice deformation. However, at the low
temperatures expected in the NPC this effect will not be present,
and direct hardening will be the dominant influence of dust (see
Sect.~\ref{sect_dust_content}).

Durham et~al.\ \cite{durham_etal_97a} propose an alternative flow law
for grain-size-independent dislocation creep, based on laboratory
creep tests at a confining pressure of 50~MPa. For the temperature
regime $T=\mbox{195--240}\;\mathrm{K}$, which corresponds
approximately to $T'=\mbox{200--245}\;\mathrm{K}$, they report the
parameters $n=4$, $p=0$,
$A_0=1.259\times{}10^{-19}\;\mathrm{s^{-1}\,Pa^{-4}}$ and
$Q=61\;\mathrm{kJ\;mol^{-1}}$.

However, for very low temperature and strain-rate conditions, as
they are expected in the polar ice caps of Mars, it is not clear
whether dislocation creep is still the predominant creep mechanism
in the polycrystalline ice aggregate. There is evidence that
other, grain-size-dependent flow mechanisms like grain-boundary
sliding become favoured instead \cite{goldsby_kohlstedt_97}.
These can be described by the parameters $n=1.8$, $p=1.4$,
$A_0=6.20\times{}10^{-14}\;\mathrm{s^{-1}\,Pa^{-1.8}\,m^{1.4}}$
and $Q=49\;\mathrm{kJ\;mol^{-1}}$ (see also \cite{nye_00}).

\begin{figure}[htb]
  \centering
  \includegraphics[scale=0.8]{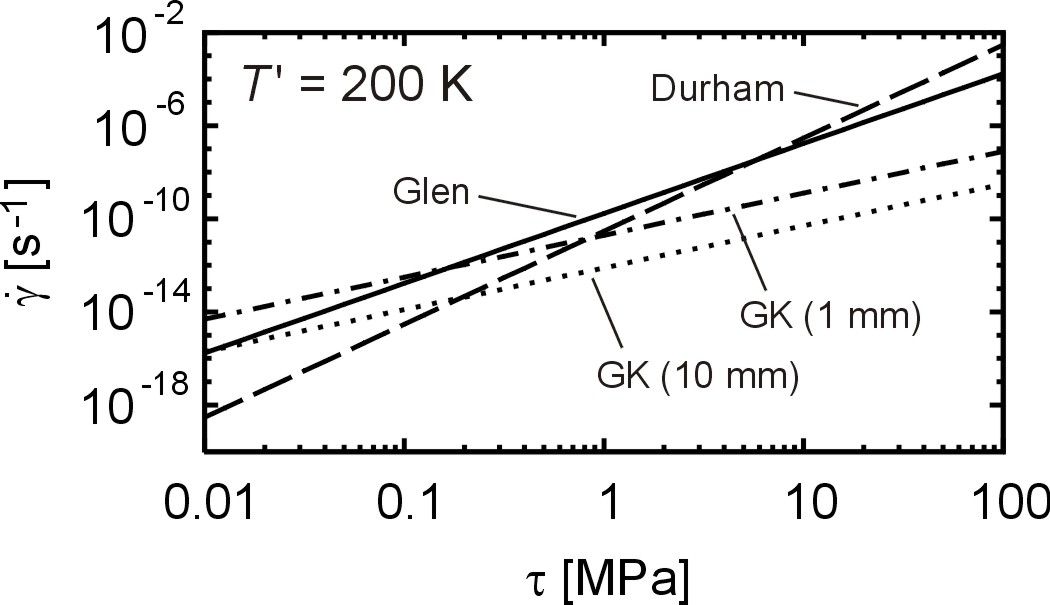}
  \caption{Shear rate $\dot{\gamma}$ vs.\ shear stress $\tau$
  for simple shear, computed by Eq.~(\ref{eq_flowlaw_simple_shear})
  for $T'=200\;\mathrm{K}$.
  Flow laws: Glen ($n=3$), Durham ($n=4$),
  GK ($n=1.8$, $p=1.4$, $d=1\;\mathrm{mm}$),
  GK ($n=1.8$, $p=1.4$, $d=10\;\mathrm{mm}$).}
  \label{fig_shear_rate}
\end{figure}

The relative contributions of the several flow laws (Glen, Durham,
Goldsby-Kohlstedt; the latter will be abbreviated as ``GK'' in the
following) can be estimated as follows. For simple shear in the
$x$-$z$ plane and $E=1$, Eq.~(\ref{eq_flowlaw_general}) reduces to
\beq
  \dot{\gamma} = 2A(T')\frac{\tau^n}{d^p},
  \label{eq_flowlaw_simple_shear}
\eeq
where $\tau=t_{xz}\dev$ is the shear stress and
$\dot{\gamma}=\partial{}v_x/\partial{}z=2D_{xz}$ is the shear rate.
Figure~\ref{fig_shear_rate} shows the shear rates resulting from
Eq.~(\ref{eq_flowlaw_simple_shear}) for $T'=200\,\mathrm{K}$ and
the stress range from 10~kPa to 100~MPa. For GK, the grain sizes
$d=1\,\mathrm{mm}$ and $10\,\mathrm{mm}$ have been assumed, based
on an estimate for the Martian north polar cap by Greve and Mahajan
\cite{greve_mahajan_05}. This range is also typical for
terrestrial ice sheets and glaciers.

It becomes evident that the relative contributions of the
different flow laws vary strongly over this stress range. For low
stresses, grain-size dependent flow with a low stress exponent
(GK) dominates, whereas for higher stresses dislocation creep with
a higher stress exponent (Glen, Durham) becomes more important.
For the shown example with $T'=200\,\mathrm{K}$, the crossover
stresses are 166.3~kPa for Glen/GK$_{d=1\,\mathrm{mm}}$, 11.3~kPa
for Glen/GK$_{d=10\,\mathrm{mm}}$, 834.0~kPa for
Durham/GK$_{d=1\,\mathrm{mm}}$ and 192.7~kPa for
Durham/GK$_{d=10\,\mathrm{mm}}$.

In order to account for the contributions of grain-size-independent
dislocation creep and other, grain-size-dependent flow mechanisms
simultaneously, a modified version of the flow law proposed by Pettit
and Waddington \cite{pettit_waddington_03} may be used. It sums up the
two contributions via
\beq
  \vecbi{D}
  = \left( E_1 A_1(T') \frac{\sigma^{n_1-1}}{d^{p_1}}
         + E_2 A_2(T')\frac{\sigma^{n_2-1}}{d^{p_2}} \right)
    \,\vecbi{t}\dev,
\eeq
where
\beq
  A_{1}(T')=(A_0)_{1}\,e^{-Q_{1}/RT'},
  \quad
  A_{2}(T')=(A_0)_{2}\,e^{-Q_{2}/RT'}.
\eeq
In this representation, the index ``1'' refers to the parameters
of either Glen's or Durham's flow law, the index ``2'' to those of
the Goldsby-Kohlstedt flow law. However, this approach will not be
further pursued here.

Note that, by introducing the \emph{viscosity} $\eta$, the flow
law (\ref{eq_flowlaw_general}) with the Arrhenius law in the form
(\ref{eq_arrhenius}) can be written in compact form as
\beq
  \vecbi{D} = \frac{1}{2\eta(T',\sigma,d)}\,\vecbi{t}\dev,
  \qquad\mbox{where}\;\,
  \eta(T',\sigma,d) = \frac{1}{2EA(T')}\frac{d^p}{\sigma^{n-1}}.
  \label{eq_flowlaw_visc}
\eeq
Similarly, the inverse flow law (\ref{eq_flowlaw_inv_general})
reads
\beq
  \vecbi{t}\dev = 2\eta(T',\delta,d)\,\vecbi{D},
  \qquad\mbox{where}\;\,
  \eta(T',\delta,d) = \frac{E_\mathrm{s}B(T')}{2}
                      \frac{d^{p/n}}{\delta^{1-1/n}}.
  \label{eq_flowlaw_inv_visc}
\eeq
Of course, the two representations of the viscosity in
(\ref{eq_flowlaw_visc}) and (\ref{eq_flowlaw_inv_visc}) are
equivalent.

Evidently, for stress exponents $n>1$, the viscosities
(\ref{eq_flowlaw_visc}), (\ref{eq_flowlaw_inv_visc})
become infinite in the low-stress/deformation limit
$\sigma,\,\delta\rightarrow{}0$, and there has been a long-standing
debate whether this is physically acceptable. Also, some experimental
evidence seems to indicate a Newtonian ($n=1$) rheology for
this regime, even though these findings have been questioned (see the
discussion by \cite{paterson_94}). There are different possibilities
of introducing a finite residual viscosity $\eta_\mathrm{res}$. The
effective stress $\sigma$ and the effective strain rate $\delta$ in
the expressions (\ref{eq_flowlaw_visc}), (\ref{eq_flowlaw_inv_visc})
can be replaced by $\max(\sigma,\,\sigma_\mathrm{res})$ and
$\max(\delta,\,\delta_\mathrm{res})$, respectively, where
$\sigma_\mathrm{res}$ and $\delta_\mathrm{res}$ are small positive
parameters. Alternatively, a small residual viscosity may be directly
added,
\beq
  \eta(T',\sigma,d) = \frac{1}{2EA(T')}\frac{d^p}{\sigma^{n-1}}
                      + \eta_\mathrm{res},
\eeq
or
\beq
  \eta(T',\delta,d) = \frac{E_\mathrm{s}B(T')}{2}
                      \frac{d^{p/n}}{\delta^{1-1/n}}
                      + \eta_\mathrm{res}.
\eeq
The disadvantage of these forms is that they are not equivalent
anymore, and an inversion according to
Eqs.~(\ref{eq_delta_sigma})-(\ref{eq_flowlaw_inv_general}) can no
longer be executed.

\subsection{Influence of liquid water}
\label{sect_water_content}

If the temperature of ice reaches the pressure melting point,
liquid water may be present within the ice matrix, which is stored
as lenses at the grain boundaries and in capillary channels at
triple grain junctions. Let $\varphi_\mathrm{w}$ be the volume
fraction of water, then the density of the ice-water mixture is
\beq
  \rho = (1-\varphi_\mathrm{w}) \rho_\mathrm{i}
         + \varphi_\mathrm{w}\rho_\mathrm{w},
  \label{eq_rho_temp_ice}
\eeq
where $\rho_\mathrm{i}=910\,\mathrm{kg\,m^{-3}}$ is the density of
pure ice and $\rho_\mathrm{w}=1000\,\mathrm{kg\,m^{-3}}$ the
density of pure water. Further, it is clear that the presence of
liquid water will reduce the viscosity.
Tobie et~al.\ \cite{tobie_etal_03}
propose an exponential softening, which can be expressed by a
stress enhancement factor $E_\mathrm{s}<1$,
\beq
  E_\mathrm{s} = e^{-b_\mathrm{w}\varphi_\mathrm{w}},
  \label{eq_es_temp_ice}
\eeq
or, equivalently, by a flow enhancement factor $E>1$,
\beq
  E = E_\mathrm{s}^{-n} = e^{nb_\mathrm{w}\varphi_\mathrm{w}},
  \label{eq_e_temp_ice}
\eeq
where $b_\mathrm{w}=45$. This value was chosen such that 5\% melt
decreases the viscosity in Eq.~(\ref{eq_flowlaw_inv_visc}) by an
order of magnitude, that is,
$E_\mathrm{s}(\varphi_\mathrm{w}\!=\!0.05)=0.1$. In terrestrial
glaciology, a linear relation reported by Paterson \cite{paterson_94}
has been widely used instead, which reads in terms of the enhancement
factors
\beq
  E = 1 + b_\mathrm{w}\varphi_\mathrm{w}
  \quad\Leftrightarrow\quad
  E_\mathrm{s}
  = E^{-1/n}
  = (1 + b_\mathrm{w}\varphi_\mathrm{w})^{-1/n},
  \label{eq_e_lin_temp_ice}
\eeq
where $b_\mathrm{w}=\frac{580}{3.2}=181.25$. This relation has
been established based on laboratory measurements for small water
contents below 1\% ($\varphi_\mathrm{w}<0.01$).

\subsection{Influence of dust}
\label{sect_dust_content}

Satellite imagery shows that parts of the Martian polar ice caps
appear dark, which indicates that they consist of ice with some amount
of mixed-in dust. Greve and Mahajan \cite{greve_mahajan_05} have laid
down that this affects the ice flow in a multiple way, in that direct
hardening can be partly compensated or even overcompensated by the
increasing density, which increases the driving stresses, and the
decreasing heat conductivity, which makes the ice at depth warmer and
therefore softer. Therefore, the average volume fraction
$\varphi_\mathrm{d}$ of dust is introduced, and the density, $\rho$,
heat conductivity, $\kappa$, and specific heat, $c$, of the ice-dust
mixture are computed as volume-fraction-weighed averages of the values
for pure ice and crustal material,
\beqa
  \rho &=& (1-\varphi_\mathrm{d}) \rho_\mathrm{i}
           + \varphi_\mathrm{d} \rho_\mathrm{c},
  \nl
  \kappa &=& (1-\varphi_\mathrm{d}) \kappa_\mathrm{i}
             + \varphi_\mathrm{d} \kappa_\mathrm{c},
  \label{eq_rho_kappa_c_mixture}
  \\
  \rho c &=& (1-\varphi_\mathrm{d}) \rho_\mathrm{i} c_\mathrm{i}
             + \varphi_\mathrm{d} \rho_\mathrm{c} c_\mathrm{c},
  \nonumber
\eeqa
with the following parameters: ice density
$\rho_\mathrm{i}=910\,\mathrm{kg\,m^{-3}}$, heat conductivity of
ice
$\kappa_\mathrm{i}=9.828\,e^{-0.0057\,T[\mathrm{K}]}\,\mathrm{W\,m^{-1}K^{-1}}$,
specific heat of ice
$c_\mathrm{i}=(146.3+7.253\,T[\mathrm{K}])\,\mathrm{J\,kg^{-1}K^{-1}}$,
density of crustal material (dust)
$\rho_\mathrm{c}=2900\;\mathrm{kg\,m^{-3}}$, heat conductivity of
crustal material (dust)
$\kappa_\mathrm{c}=2.5\;\mathrm{W\,m^{-1}\,K^{-1}}$, specific heat
of crustal material (dust)
$c_\mathrm{c}=1000\,\mathrm{J\,kg^{-1}K^{-1}}$. The additional
density factors in Eq.~(\ref{eq_rho_kappa_c_mixture})$_3$ are
necessary because the averaging procedure requires volumetric
quantities, whereas the specific heat is taken per mass unit.

Direct hardening is described by a stress enhancement factor
$E_\mathrm{s}>1$ based on laboratory measurements of the
deformation of ice-dust compounds,
\beq
  E_\mathrm{s} = e^{b_\mathrm{d}\varphi_\mathrm{d}},
\eeq
where $b_\mathrm{d}=2$ and $\varphi_\mathrm{d}\le{}0.56$
\cite{durham_etal_97a}. This is equivalent to a flow enhancement
factor
\beq
  E = E_\mathrm{s}^{-n} = e^{-nb_\mathrm{d}\varphi_\mathrm{d}}.
  \label{eq_enh_mixture}
\eeq
Hence, for given stress, temperature and grain-size conditions and
a stress exponent $n=3$ a dust content of 10\%
($\varphi_\mathrm{d}=0.1$) leads to an almost twice as hard
material ($E=0.55$) compared to pure ice.

\section{Model equations for ice flow}
\label{sect_model_ice_flow}

The thermo-mechanical problem of ice flow in a planetary
environment can be described by the balance equations of mass,
momentum and energy (e.g.\ \cite{hutter_83}). For a
density-preserving (incompressible) medium, which holds for ice in
good approximation despite the variability expressed by
Eqs.~(\ref{eq_rho_temp_ice}) and
(\ref{eq_rho_kappa_c_mixture})$_1$, the mass balance (continuity
equation) reads
\beq
   \divg\vecbi{v} = 0.
  \label{eq_mass_balance}
\eeq
For evolving ice bodies, it is convenient to vertically integrate
the continuity equation. This yields the ice-thickness ($H$)
equation
\beq
  \pabl{H}{t} = -\divg\vecbi{Q} + a_\mathrm{s} - a_\mathrm{b},
  \label{eq_ice_thickness}
\eeq
where $\vecbi{Q}$ is the volume flux (vertically integrated
horizontal velocity), and $a_\mathrm{s}$ and $a_\mathrm{b}$ are
the mass balances at the surface (positive for supply) and the
bottom (positive for loss), respectively. The momentum balance
yields with the flow law (\ref{eq_flowlaw_inv_visc}) the
Stokes equation
\beq
  - \grad P
  + \divg\!\!\left[\eta\left(\grad\vecbi{v}+(\grad\vecbi{v})\tra\right)\right]
  + \rho\vecbi{g} = \mathbf{0}
  \label{eq_stokes}
\eeq
($\vecbi{g}$: vectorial gravity acceleration), in which the
acceleration term $\rho\,\D\vecbi{v}/\D{}t$ has been neglected due
to the very low flow velocities to be expected. From the energy
balance, Fourier's law of heat conduction
\beq
  \vecbi{q} = -\kappa(T)\,\grad{}T
  \label{eq_fourier_heat_conduction}
\eeq
($\vecbi{q}$: heat flux, $\kappa$: heat conductivity) and the
caloric equation of state
\beq
  u = \int\limits_{T_0}^{T}c(\bar{T})\,\D\bar{T}
  \label{eq_int_energy_temp}
\eeq
($u$: specific internal energy, $c$: specific heat), the
temperature-evolution equation
\beq
  \rho c \left( \pabl{T}{t} + \vecbi{v}\cdot\grad{}T \right)
  = \divg(\kappa\,\grad{}T) + 4\eta\,\delta^2 + r
  \label{eq_ice_flow_temperature}
\eeq
results. In this relation, the production term $4\eta\,\delta^2$
is the strain heating, and the source term $r$ denotes the
volumetric heating due to radiation and tidal dissipation.

The above equations need to be complemented by dynamic and
thermodynamic boundary conditions at the surface and the bottom of
the respective ice body. If we assume that the surface is in
contact with the atmosphere, then it can be described in good
approximation as stress-free, that is,
\beq
  \left.\vecbi{t}\cdot\vecbi{n}\right|_\mathrm{s} = \mathbf{0}
\eeq
(where $\vecbi{n}$ is the outer normal unit vector, and the subscript
``s'' denotes the surface). The surface temperature $T_\mathrm{s}$
can be prescribed directly as a Dirichlet condition.

If the bottom is a rigid ice/rock, ice/regolith or ice/sediment
interface, no-slip conditions can be employed,
\beq
  \vecbi{v}_\mathrm{b} = \mathbf{0}
\eeq
(the subscript ``b'' stands for the bottom). As for the temperature
field, let us assume that the basal heat flux into the ice,
$\vecbi{q}_\mathrm{b}$, is known. This yields the Neumann condition
\beq
  \left.\kappa\pabl{T}{\vecbi{n}}\right|_\mathrm{b} = \vecbi{q}_\mathrm{b},
\eeq
where $\vecbi{n}$ is again the outer normal unit vector.

The situation is different if the bottom is an ice/water interface.
In this case, the basal stress conditions are governed by the
hydrostatic pressure $P_\mathrm{b}$ of the water at the interface,
\beq
  \left.\vecbi{t}\cdot\vecbi{n}\right|_\mathrm{b}
  = -P_\mathrm{b}\,\vecbi{n},
\eeq
and the bottom temperature equals the pressure melting point,
\beq
   T_\mathrm{b} = T_\mathrm{m}.
\eeq
Provided that the role of impurities is negligible, $T_\mathrm{m}$ can
be obtained from Eq.~(\ref{eq_p_melt_temp}).

\section{Polar caps of Mars}
\label{sect_mpc}

\subsection{Large-scale simulations}
\label{sect_mpc_sico}

The dynamic and thermodynamic state of the present-day polar caps
of Mars will now be simulated with the ice-sheet model SICOPOLIS
(``SImulation COde for POLythermal Ice Sheets''). This model was
developed in the mid-1990's for terrestrial applications and has
later been adapted to the north-polar cap of Mars (see
\cite{greve_mahajan_05}, and references therein).  It solves the
ice-flow equations described in Sect.~\ref{sect_model_ice_flow}
based on the shallow-ice approximation \cite{hutter_83,
morland_84}, that is, the flow regime is assumed to be simple,
bed-parallel shear, the pressure is hydrostatic and lateral shear
stresses as well as normal stress deviators are neglected (see
also \cite[this volume]{calov_06}). Inputs from the environment
are specified by the mean annual surface temperature, the net
surface mass balance (ice accumulation minus ablation) and the
basal (``geothermal'') heat flux from the underlying lithosphere.
The numerical solution of the model equations is carried out by a
finite-difference integration technique.

\begin{figure}[htb]
  \centering
  \includegraphics[scale=1.2]{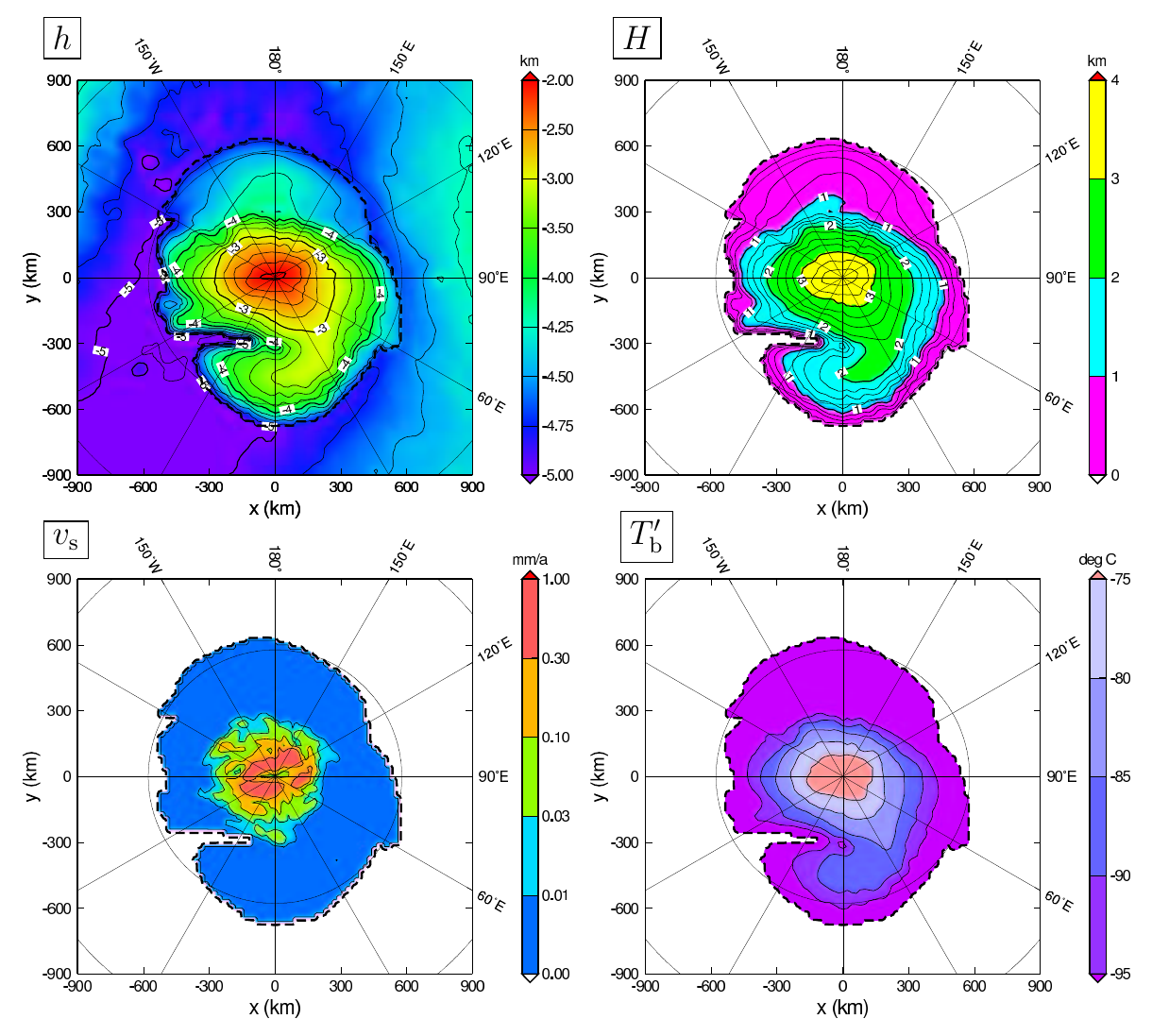}
  \caption{North-polar cap of Mars:
  MOLA surface topography $h$ \cite{smith_etal_99a},
  computed ice thickness $H$, computed surface velocity $v_\mathrm{s}$
  and computed homologous basal temperature $T'_\mathrm{b}$ for
  present-day steady-state conditions. Heavy-dashed lines indicate the
  ice-cap margin.}
  \label{fig_nmars}
\end{figure}

\begin{figure}[htb]
  \centering
  \includegraphics[scale=1.2]{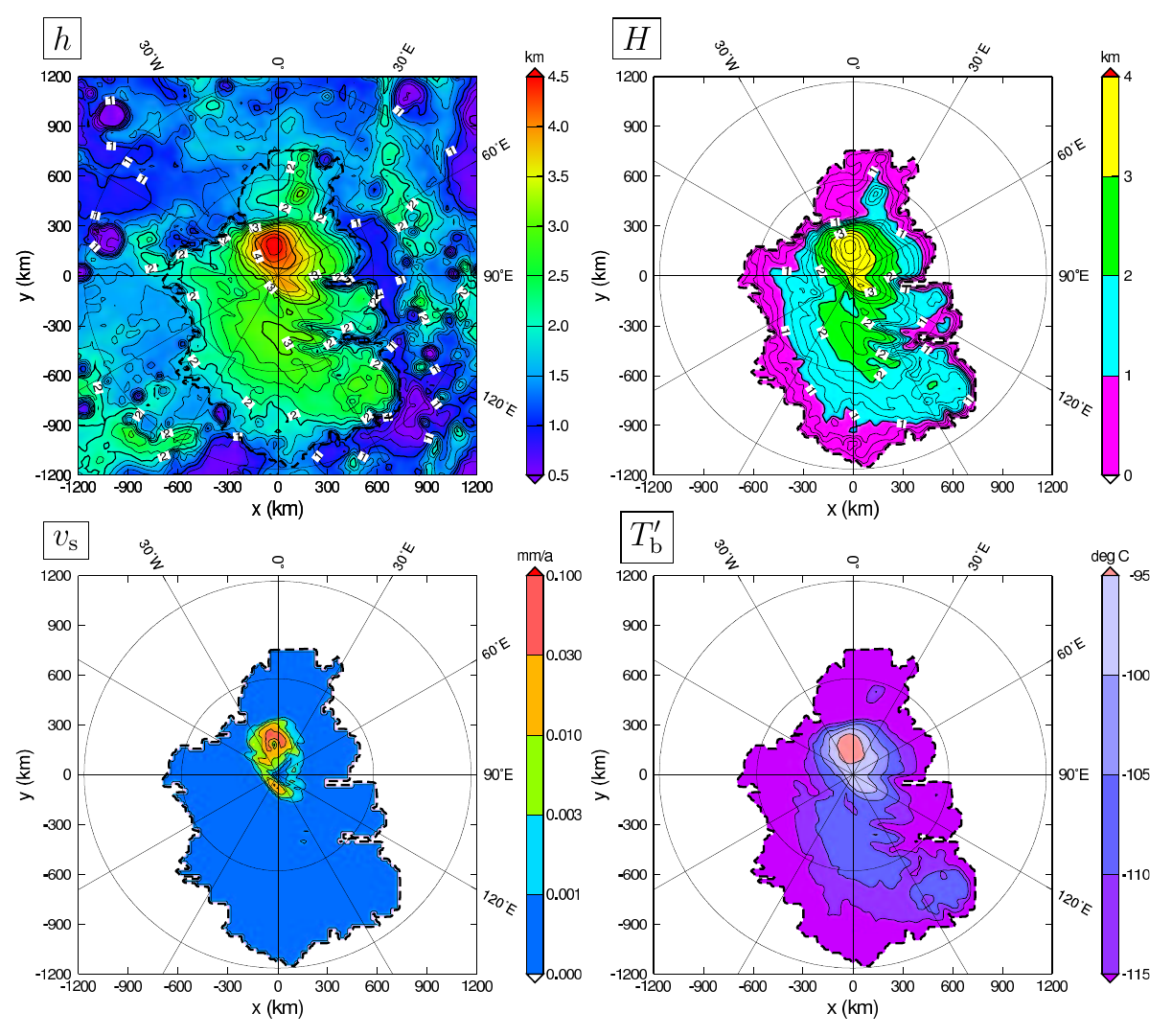}
  \caption{South-polar cap of Mars:
  MOLA surface topography $h$ \cite{smith_etal_99a},
  computed ice thickness $H$, computed surface velocity $v_\mathrm{s}$
  and computed homologous basal temperature $T'_\mathrm{b}$ for
  present-day steady-state conditions. Heavy-dashed lines indicate the
  ice-cap margin.}
  \label{fig_smars}
\end{figure}

Here, we only consider steady-state conditions, that is, the ice
surface is held fixed as given by the Mars Orbiter Laser Altimeter
(MOLA) data of the Mars Global Surveyor (MGS) spacecraft
\cite{smith_etal_99a, zuber_etal_98}. This makes it unnecessary
to prescribe the surface mass balance. The topographies are shown
in Figs.~\ref{fig_nmars} and \ref{fig_smars} (top left panels) for
the north- and south-polar caps and their surroundings. It is
striking that the overall shape of the north-polar cap is quite
regular and smooth, whereas the south-polar cap appears much
more rugged.

The ice temperature, velocity and bottom topography are allowed to
evolve freely until the steady state is reached. Surface
temperature is delivered by the model by Grieger (pers.\ comm.\
2004; see also http://www.space-vision.biz/marstemperatures.html),
which is based on a zonal and daily mean energy balance including
a simple scheme for CO$_2$ condensation and evaporation. Results
agree well with those of the Martian Climate Database
\cite{lewis_etal_99}; however, both methods fail in reproducing
the observed year-round CO$_2$ cover of the southern residual cap.
Therefore, the southern mean-annual surface temperature is
corrected such that it equals the CO$_2$ sublimation temperature
of $-128\degC$ within $85\degS$. Further, the geothermal heat flux
is assumed to be $35\,\mathrm{mW\,m^{-2}}$
\cite{schubert_etal_92}, and the source term $r$ in the
temperature evolution equation (\ref{eq_ice_flow_temperature}) is
neglected.

The unknown topography of the solid ground below the polar caps is
computed in two steps. First, the equilibrated ground for ice-free
conditions, which is required as a reference topography, is determined
by a smooth extrapolation of the ice-free ground surrounding the north
and south polar cap, respectively \cite{greve_etal_04a}. Second, the
actual ground topography is obtained by superposing the isostatic
deflection (downward displacement $w$) of the underlying lithosphere
due to the ice load. By modelling the lithosphere as a thin elastic
plate, the isostatic deflection is governed by the bi-potential
equation
\beq
  K_\mathrm{l}\,\nabla^4{}w = \rho g H - \rho_\mathrm{a} g w,
  \label{eq_elast_litho}
\eeq
where $K_\mathrm{l}=10^{25}\,\mathrm{N\,m}$ is the flexural stiffness
of the lithosphere, $\nabla^4$ the bi-potential operator in the
horizontal plane and $\rho_\mathrm{a}=3500\,\mathrm{kg\,m^{-3}}$ the
density of the asthenosphere (viscous mantle layer below the elastic
lithosphere). The two load terms on the right-hand side are the ice
load itself ($\rho{}gH$) and the counteracting buoyancy force which
the deflected lithosphere experiences from the asthenosphere below
($\rho_\mathrm{a}gw$). Note that the ice thickness $H$ depends on $w$
and is therefore part of the solution. In transient scenarios, which
are not considered here, the elastic deflection (\ref{eq_elast_litho})
is not assumed instantaneously, so that an additional evolution
equation is required for the non-equilibrium displacement
\cite{greve_01, lemeur_huy_96}.

The simulations have been carried out by applying Glen's flow law
as discussed in Sect.~\ref{sect_ice_flow}. The dust content is
assumed to be 20\% ($\varphi_\mathrm{d}=0.2$, see
Sect.~\ref{sect_dust_content}). Horizontal resolution is 20~km in
the stereographic plane with standard parallel $71\degN$/S,
vertical resolution is 51 grid points in the cold-ice column and
11 grid points in the lithosphere column, and the time-step is
1000 years.

Results for the ice thickness, the surface velocity and the
homologous basal temperature are shown in Figs.~\ref{fig_nmars}
and \ref{fig_smars} (top right and bottom panels) for both polar
caps. The north-polar cap assumes its maximum thickness of 3.62~km
almost exactly at the pole, whereas the maximum thickness of the
south-polar cap, 3.88~km, is offset by approx.\ 150~km in
$10\degW$ direction. Of the entire ice volume of the NPC,
which is $1.53\times{}10^6\,\mathrm{km}^3$, a fraction of
$1.23\times{}10^6\,\mathrm{km}^3$ (80\%) is above the level of the
equilibrated ground, and the remaining 20\% are due to the
isostatic deflection of the lithosphere. This ratio is almost the
same for the SPC, with a total volume of
$2.30\times{}10^6\,\mathrm{km}^3$ and a volume above the
equilibrated ground of $1.85\times{}10^6\,\mathrm{km}^3$.

Compared to terrestrial ice sheets and glaciers with typical flow
velocities of tens to hundreds of meters per year, the flow of the
Martian polar caps is very slow. The NPC reaches a maximum surface
velocity of $0.98\,\mathrm{mm\,a^{-1}}$, and south of $85\degN$
surface velocities are everywhere less than
$0.01\,\mathrm{mm\,a^{-1}}$. Therefore, the active, dynamic zone
only consists of the interior, thick part of the ice cap. The
situation is even more extreme for the SPC, where the maximum
surface velocity is as small as $0.055\,\mathrm{mm\,a^{-1}}$, and
by far the largest part of the ice cap flows at speeds slower than
a micrometer per year (note the different scales of the colour
bars). Therefore, the SPC can be considered as essentially
stagnant.

Note that the 20-km resolution applied here does not resolve
small-scale structures like the scarp/trough systems, which may
lead to locally enhanced flow velocities. This will be discussed
below (Sect.~\ref{sect_mpc_scarp}) for the NPC.

The slow flow velocities of the Martian polar caps lead to a
conduction-dominated heat transport within the ice. The ratio
of convection to conduction is given by the \emph{Peclet number}
\beq
  P\!e = \frac{[U][L]}{[\alpha]},
\eeq
where $[U]$ is a velocity scale, $[L]$ is a scale of extent and
$[\alpha]$ is a scale for the thermal diffusivity
$\alpha=\kappa/(\rho{}c)$. With $[U]=0.1\,\mathrm{mm\,a^{-1}}$ for
the NPC and $0.01\,\mathrm{mm\,a^{-1}}$ for the SPC, respectively,
$[L]=300\,\mathrm{km}$ (only the inner, dynamic region considered;
see velocity panels of Figs.~\ref{fig_nmars} and \ref{fig_smars})
and $[\alpha]=2\times{}10^{-6}\,\mathrm{m^2\,s^{-1}}$ (for
$-100\degC$ and 20\% dust), one obtains Peclet numbers of
$P\!e\approx{}0.5$ for the NPC and $P\!e\approx{}0.05$ for the
SPC, respectively. Evidently, both values are less than unity, so
that conduction outweighs convection. This leads to temperature
profiles which increase essentially linearly with depth (not
shown). By contrast, for terrestrial ice sheets and ice caps,
$P\!e\gg{}1$ always holds, so that their temperature fields are
mainly governed by heat convection.

The different flow behaviour of the NPC and SPC is mainly due to
the lower ice temperatures of the SPC, which are a consequence of
the $\sim{}20\degC$ difference in surface temperatures. Since the
thicknesses of both caps are similar, and the temperature
distribution is mainly controlled by heat conduction, this
transfers directly to a $\sim{}20\degC$ difference in average
basal temperatures (again, note the different scales of the colour
bars in the respective panels of Figs.~\ref{fig_nmars} and
\ref{fig_smars}). The highest homologous basal temperatures are
$-69.1\degC$ for the NPC and $-89.7\degC$ for the SPC,
respectively, which demonstrates the insulating effect of the ice
caps against the much colder surface temperatures. Nevertheless,
the basal temperatures are in any case far below the pressure
melting point. This is a very robust result, so that the presence
of large amounts of subglacial liquid water as a potential habitat
for Martian lifeforms can essentially be ruled out.

The simulation for the north-polar cap has been re-run with the
alternative flow laws shown in Fig.~\ref{fig_shear_rate} (GK with
$d=1\,\mathrm{mm}$ and $10\,\mathrm{mm}$, Durham). As it was already
found by Greve and Mahajan \cite{greve_mahajan_05}, this has a very
significant influence on the computed flow velocities. The maximum
surface velocity varies by almost three orders of magnitude, from the
largest value $3.02\,\mathrm{mm\,a^{-1}}$ for GK$_{d=1\,\mathrm{mm}}$
via the above-mentioned $0.98\,\mathrm{mm\,a^{-1}}$ for the reference
simulation (Glen) and $0.12\,\mathrm{mm\,a^{-1}}$ for
GK$_{d=10\,\mathrm{mm}}$ to only $0.013\,\mathrm{mm\,a^{-1}}$ for
Durham. As expected, this order corresponds to the low-stress regime
in Fig.~\ref{fig_shear_rate}. Since the creep mechanism is probably a
combination of dislocation creep (Glen, Durham) and
grain-size-dependent creep (GK), we consider a value of the order of
$1\,\mathrm{mm\,a^{-1}}$ as most likely. By contrast, the basal
temperature is virtually unaffected by the assumed flow law, which is
a consequence of the dominance of flow-independent heat conduction
over flow-dependent heat convection.

\subsection{Detailed simulations of the scarps and troughs}
\label{sect_mpc_scarp}

In the large-scale simulations discussed above
(Sect.~\ref{sect_mpc_sico}), the spiralling scarps and troughs
which cut up to several hundred meters into the surface of the
Martian polar caps, have not been resolved. This was tackled in
a study by Hvidberg \cite{hvidberg_03}, where a flowline of the
north-polar cap extending from the pole in $160\degE$ direction
was investigated in detail, and an axisymmetric ice cap was
assumed. The flowline was discretized by a finite-element grid at
kilometer-scale horizontal resolution, and the ice-flow equations
of Sect.~\ref{sect_model_ice_flow} were solved without further
approximations. Otherwise, the model set-up is very similar to
that described in Sect.~\ref{sect_mpc_sico}.

\begin{figure}[htb]
  \centering
  \includegraphics[scale=0.94]{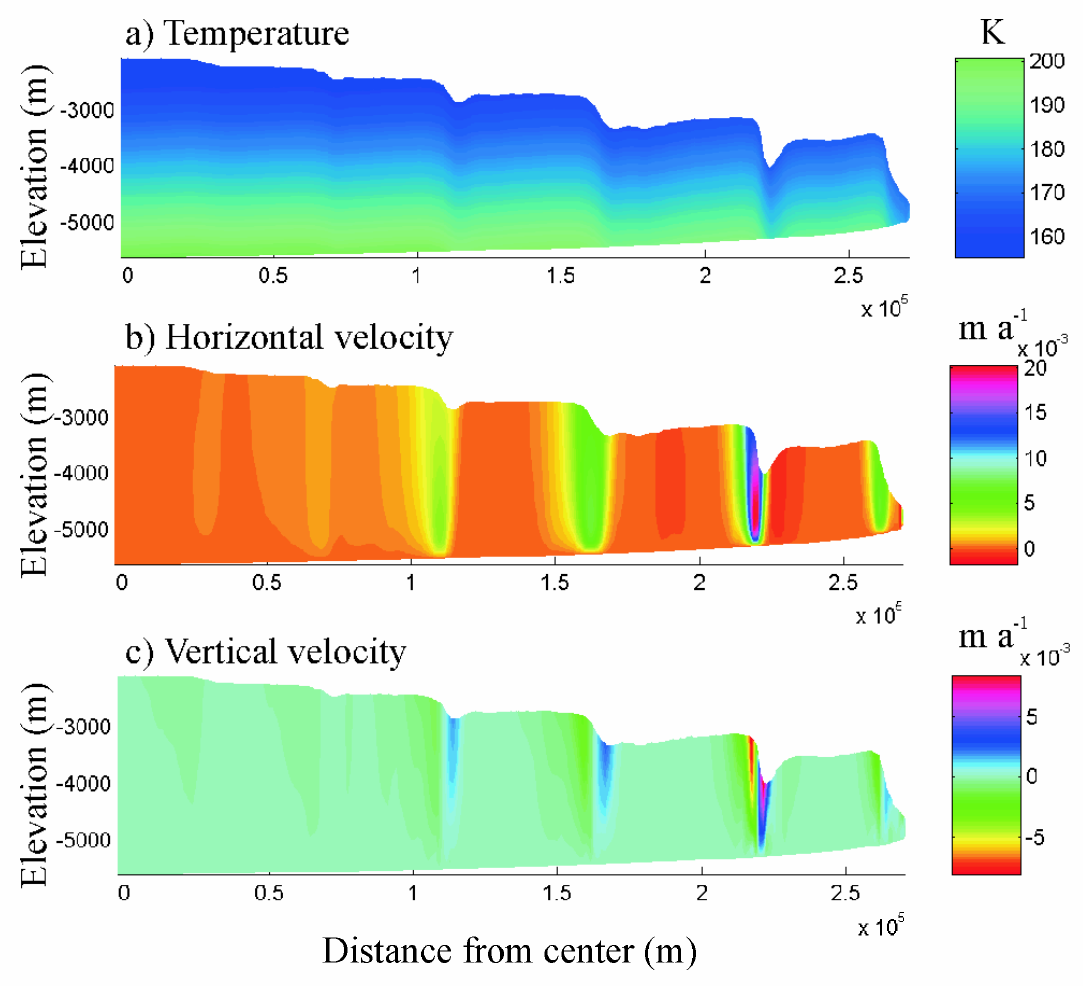}
  \caption{Computed temperature and velocity field of the north
  polar cap of Mars, for a flowline extending southward from the pole in
  $160\degE$ direction.
  Figure by Hvidberg \cite[her Fig.~2]{hvidberg_03}.}
  \label{fig_hvidberg_fig2}
\end{figure}

\begin{figure}[htb]
  \centering
  \includegraphics[scale=0.94]{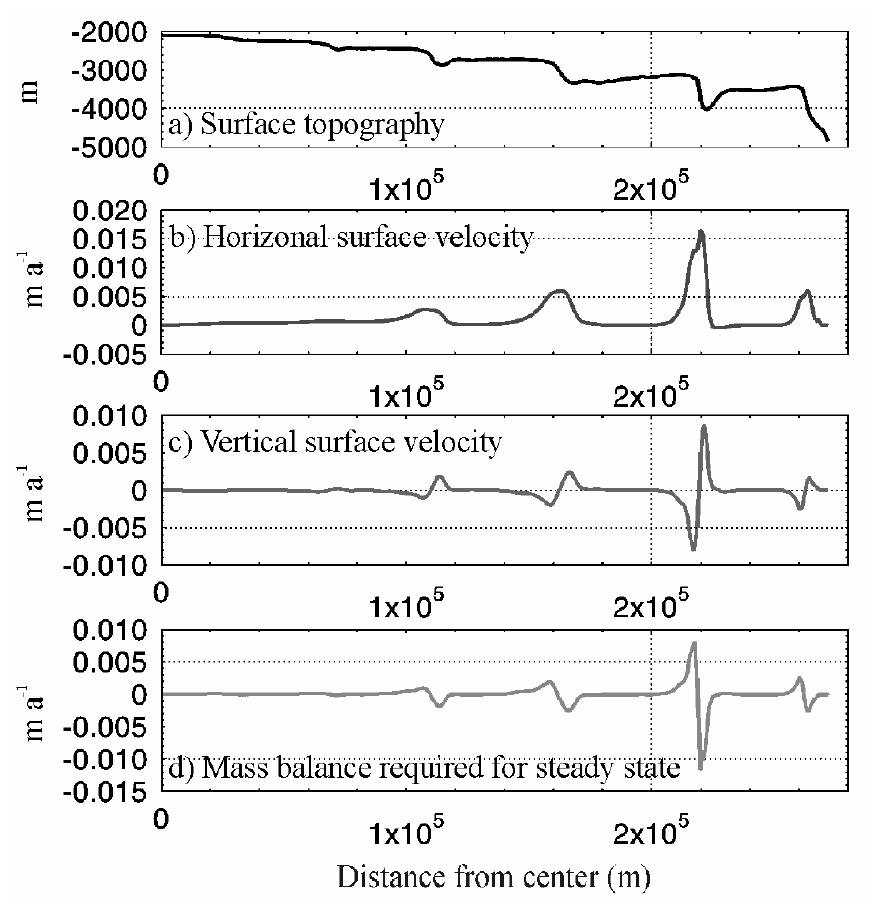}
  \caption{MOLA surface topography, computed surface velocity and computed
  steady-state mass balance corresponding to the simulation shown in
  Fig.~\ref{fig_hvidberg_fig2}.
  Figure by Hvidberg \cite[her Fig.~3]{hvidberg_03}.}
  \label{fig_hvidberg_fig3}
\end{figure}

The temperature and flow fields computed by Hvidberg
\cite{hvidberg_03} are shown in Fig.~\ref{fig_hvidberg_fig2}. While
the overall values agree very well with those reported above for the
large-scale simulations (Sect.~\ref{sect_mpc_sico},
Fig.~\ref{fig_nmars}), it becomes evident that the topographic
disturbances imposed by the scarps/troughs result in significant
disturbances of the ice flow which propagate all the way down to the
bottom. The local flow accelerations are highlighted in
Fig.~\ref{fig_hvidberg_fig3}, which shows the surface velocities and
the surface mass balance required for maintaining the steady state.
For the most pronounced trough about 220~km away from the pole, the
horizontal surface velocity is larger than $15\,\mathrm{mm\,a^{-1}}$,
and even the vertical velocity reaches values of almost
$\pm{}10\,\mathrm{mm\,a^{-1}}$, distributed such that the trough would
close in the absence of any mass exchange.  In order to keep the
troughs open, this must be balanced by a mass exchange at the surface
such that ice accumulates outside the troughs, but is removed from
within them at a rate of some millimeters per year
(Fig.~\ref{fig_hvidberg_fig3}, bottom panel).  This exchange pattern
was already proposed by Fisher \cite{fisher_93, fisher_00} and termed
``accublation model''.  Physical processes behind the ``accublation''
exchange are likely differential ablation due to the albedo contrast
(white material outside vs.\ darker material inside the troughs)
and/or enhanced wind erosion due to the formation of local turbulences
in the troughs.

\clearpage

\section{Icy shell of Europa}
\label{sect_eur_is}

While the polar caps of Mars are localized ice masses on the
planet's surface which rest on solid land, the icy shell of Europa
envelops the entire planetary body and is underlain most likely by
a deep ocean. Also, the supposed thickness within the range of
10--50~km is much larger than that of the Martian polar caps. This
allows the formation of convection cells, so that vertical motion
may play a much larger role.

\begin{figure}[htb]
  \centering
  \includegraphics[scale=0.95]{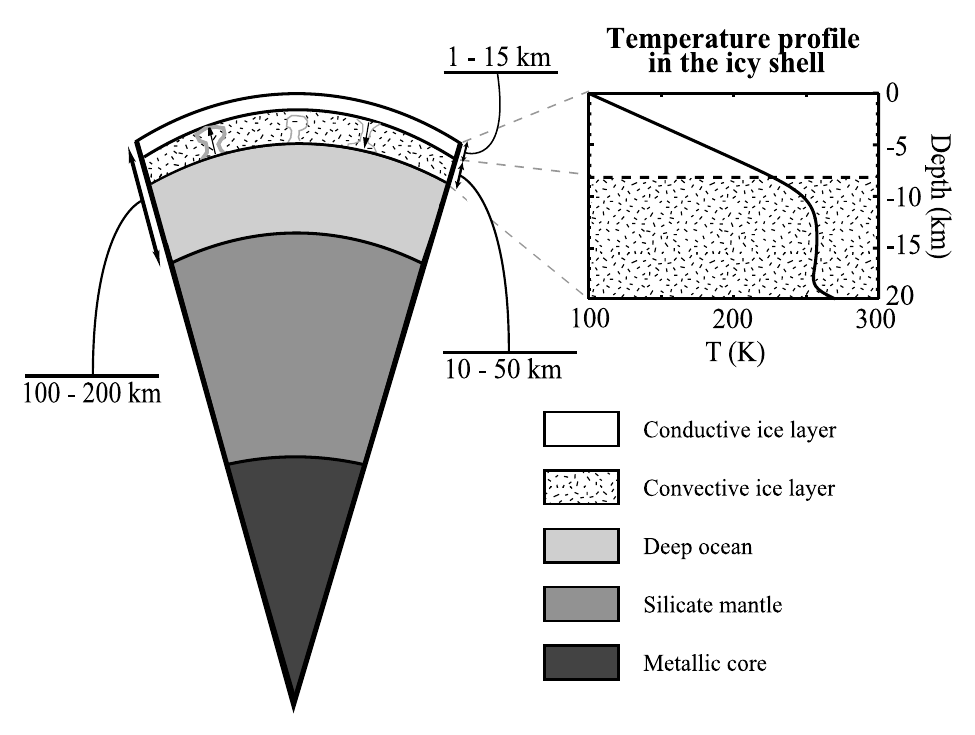}
  \caption{Model of the radial structure of Europa with a 20-km
  thick outer ice shell underlain by a deep ocean. The ice shell
  consists of an upper conductive lid and a lower convective
  layer. A possible temperature profile is shown in the inset plot.
  Figure by Tobie et~al.\ \cite[their Fig.~1]{tobie_etal_03}.}
  \label{fig_tobie_etal_fig1}
\end{figure}

Tobie et~al.\ \cite{tobie_etal_03} investigated this problem by
assuming a 20-km thick ice shell with a surface temperature of 100~K
and a bottom temperature of 270~K, which is approximately equal to the
pressure melting point (Fig.~\ref{fig_tobie_etal_fig1}). The ice-flow
equations of Sect.~\ref{sect_model_ice_flow} are solved by finite
differences in a two-dimensional cross-section of 40~km width, and the
domain is treated as Cartesian. For the viscosity law, a simplified
Newtonian viscosity is assumed, that is, $n=1$ and $p=0$ [see
Eqs.~(\ref{eq_flowlaw_visc}) and (\ref{eq_flowlaw_inv_visc})], and the
dependence on the homologous temperature $T'$ is modelled by an
activation energy of $Q=50\;\mathrm{kJ\;mol^{-1}}$, which is close to
the value of Goldsby and Kohlstedt \cite{goldsby_kohlstedt_97}. The
bottom viscosity is set to $1.5\times{}10^{14}\,\mathrm{Pa\,s}$ in the
reference simulation.  The tidal dissipation $r$ in the energy balance
(\ref{eq_ice_flow_temperature}) is calculated by assuming a
viscoelastic response of Europa to the tidal forcing, and the
parameters of the Maxwellian rheology are prescribed variably in the
four layers indicated in Fig.~\ref{fig_tobie_etal_fig1}. This can lead
to partial melting in the lower parts of the ice shell, the effect of
which is accounted for by
Eqs.~(\ref{eq_rho_temp_ice})--(\ref{eq_e_temp_ice}). In the vertical
component of the Stokes equation (\ref{eq_stokes}), the buoyancy force
due the varying ice density, which results from partial melting and
thermal expansion, is added.

\begin{figure}[htb]
  \centering
  \includegraphics[scale=0.95]{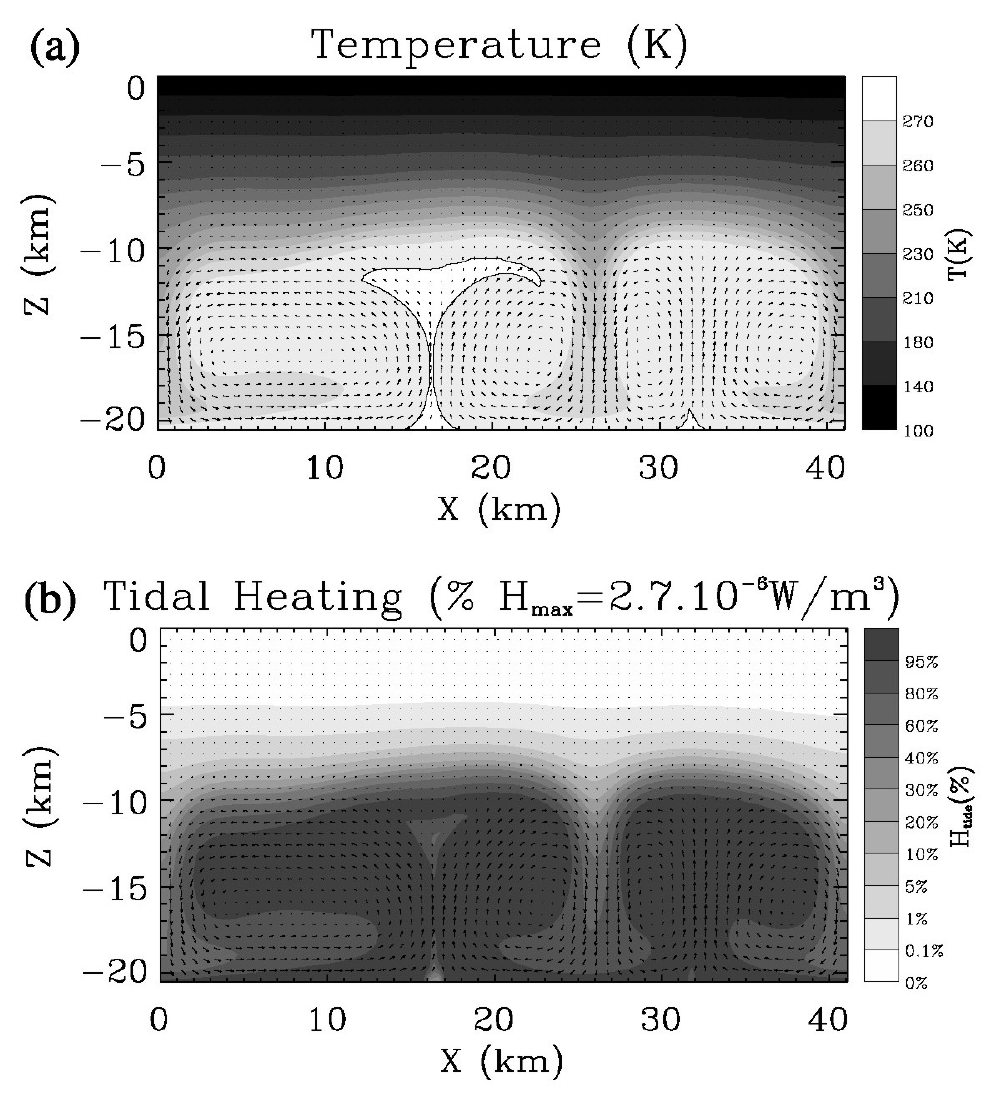}
  \caption{Computed flow field (arrows in both panels),
  temperature (panel a) and tidal heating (panel b) in
  Europa's ice shell for the reference simulation by
  Tobie et~al.\ \cite[their Fig.~4]{tobie_etal_03}.
  The solid contour in panel (a) delimits the area where the temperature
  is at the pressure melting point.}
  \label{fig_tobie_etal_fig4}
\end{figure}

The resulting fields of ice flow, temperature and tidal heating are
shown in Fig.~\ref{fig_tobie_etal_fig4}. As already sketched in
Fig.~\ref{fig_tobie_etal_fig1}, the huge temperature (and therefore
viscosity) differences lead to the formation of an upper, conductive
sublayer of approximately 8~km thickness, which is essentially rigid,
and in which the temperature increases linearly with depth. By
contrast, in the underlying convective sublayer significant ice flow
occurs in the form of convection cells, with a maximum horizontal
velocity of $0.29\,\mathrm{m\,a^{-1}}$, a maximum upward velocity of
$0.265\,\mathrm{m\,a^{-1}}$ and a maximum downward velocity of
$0.352\,\mathrm{m\,a^{-1}}$. Therefore, vertical temperature
gradients are small in the convective sublayer. In the rising plumes,
the ice temperature reaches the pressure melting point, and
consequently partial melting occurs. The tidal heating is negligible
in the upper, conductive sublayer, but reaches significant values up
to $2.7\times{}10^{-6}\,\mathrm{W\,m^{-3}}$ in the lower parts of the
ice shell. It is also interesting to note that the convection cells
entail large lateral gradients of the temperature and the tidal
heating.

\begin{figure}[htb]
  \centering
  \includegraphics[scale=1.0]{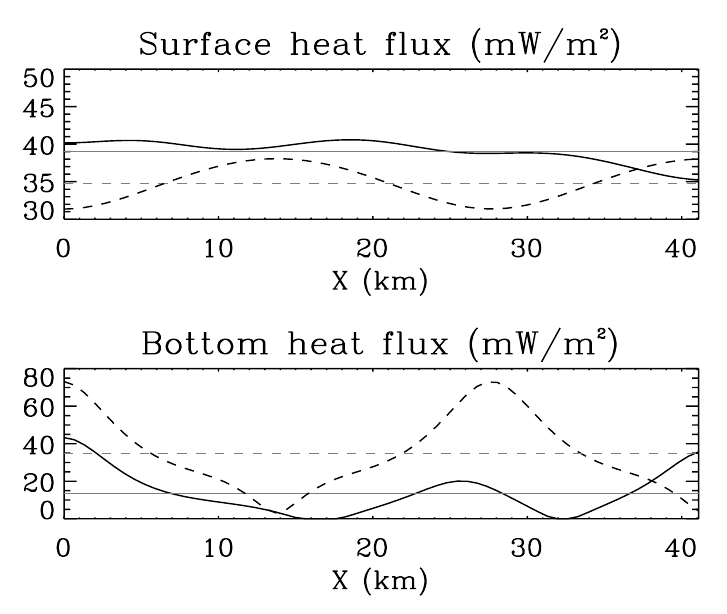}
  \caption{Surface and bottom heat flux corresponding to the
  simulation shown in Fig.~\ref{fig_tobie_etal_fig4} (solid line).
  The dashed line is for the same simulation without tidal
  heating \cite[their Fig.~6]{tobie_etal_03}.}
  \label{fig_tobie_etal_fig6}
\end{figure}

Fig.~\ref{fig_tobie_etal_fig6} shows the corresponding heat flux
at the surface and the bottom of the ice shell (solid line in the
plots). As it can be expected, the lateral variability of the heat
flux at the surface is rather small, and the average value is
approximately $40\,\mathrm{mW\,m^{-2}}$. By contrast, at the
bottom, the convection cells produce a heat-flux pattern highly
variable in space and time (the latter is not shown in the
snapshot figure) with an average of about one fourth of the
surface heat flux.

In addition to the reference simulation discussed above, Tobie et~al.\
\cite{tobie_etal_03} studied also the effects of different bottom
viscosities, tidal dissipation rates (dashed line in
Fig.~\ref{fig_tobie_etal_fig6}, etc.) and ice-shell thicknesses.  This
shall not be further reported here.

\section{Concluding remarks}
\label{sect_conclusion}

We have seen that water ice plays an important role in the climate
systems of Earth and Mars. Further, it is a major constituent of the
bulk volumes of the satellites of Jupiter, Saturn, Uranus and Neptun
as well as the planet Pluto. The ice sheets, ice caps and glaciers on
Earth and Mars show glacial flow, driven by their own weight. By
contrast, ice flow driven by thermal convection can occur in the crust
and interior of the icy satellites, depending on the temperature
gradient between the warm interior and the cold surface. Here, we have
limited the discussions to the ``ordinary'' ice~Ih. However, at
greater depths the occurrence of high-pressure phases can be expected,
the rheology of which is poorly known.

Further ``ices'' made up of other moderately volatile substances are
known or supposed to occur, for example, on Mars (CO$_2$), Io
(SO$_2$), Europa, Ganymede and Callisto (NH$_3$), Titan (CH$_4$ and
other carbohydrates), Triton and Pluto (N$_2$, CH$_4$).  However, from
a fluid-dynamical point of view, very little is known about their
relevance, and it has not been possible to assess so far whether these
materials are significant parts of flowing ice masses somewhere in the
solar system.

\section*{Acknowledgements}

The author wishes to thank K.~Hutter for his kind invitation to
contribute this article to the current issue of the
\emph{GAMM-Mitteilungen}. Comments by R.~Calov and K.~Hutter
on the draft version of the manuscript are gratefully
acknowledged.
Figure~\ref{fig_ice_phases} reproduced by permission of Oxford University
Press.
Figure~\ref{fig_iceIh} reprinted with permission from Elsevier.
Figures~\ref{fig_hvidberg_fig2} and \ref{fig_hvidberg_fig3} reprinted
from the \emph{Annals of Glaciology} with
permission of the International Glaciological Society.
Figures~\ref{fig_tobie_etal_fig1}, \ref{fig_tobie_etal_fig4} and
\ref{fig_tobie_etal_fig6} reproduced by permission of the American
Geophysical Union.



\begin{appendix}

\section{Pressure dependence of ice flow}
\label{sect_pressure}

The flow rate factor (\ref{eq_arrhenius}) can be written as
\beqa
  A(T')
  &=& A_0\exp\left(-\frac{Q}{RT'}\right)
  \nl
  &=& A_0\exp\left(-\frac{Q}{R(T+\beta{}P)}\right)
   =  A_0\exp\left(-\frac{Q}{RT(1+\beta{}P/T)}\right),
\eeqa
where the linearized melting-point depression
(\ref{eq_melt_linear}) has been used for simplicity. For the
stability range of ice~Ih, $P\apprle{}200\;\mathrm{MPa}$, we have
$\beta{}P<{}20\;\mathrm{K}$, and therefore $\beta{}P/T\ll{}1$.
This allows the Taylor approximation
\beq
  A(T')
  = A_0\exp\left(-\frac{Q}{RT}\Big(1-\frac{\beta{}P}{T}\Big)\right)
  = A_0\exp\left(-\frac{1}{RT}\Big(Q-\frac{Q\beta{}P}{T}\Big)\right).
\eeq
By comparing this with the flow rate factor in the general
form (\ref{eq_arrhenius_general}), the activation volume which
corresponds to the simplified form (\ref{eq_arrhenius}) is
\beq
  PV = -\frac{Q\beta{}P}{T}
  \quad\Rightarrow\quad
  V = -\frac{Q\beta{}}{T}.
\eeq
For $Q=60\;\mathrm{kJ\;mol^{-1}}$ and $T=200\;\mathrm{K}$,
this yields $V=-2.94\times{}10^{-5}\;\mathrm{m^3\,mol^{-1}}$,
which agrees well with the (quite uncertain) values reported by
Paterson \cite{paterson_94},
$V=-1.7\times{}10^{-5}\;\mathrm{m^3\,mol^{-1}}$, and by
Durham et~al.\ \cite{durham_etal_97a},
$V=-1.3\times{}10^{-5}\;\mathrm{m^3\,mol^{-1}}$. Therefore, the
pressure dependence of the flow rate factor is described
reasonably well by the simplified form (\ref{eq_arrhenius}) which
depends on the homologous temperature only.

\end{appendix}

\end{document}